\documentclass{article}

\usepackage[nonatbib,final]{neurips_2021_ml4ps}

\usepackage[utf8]{inputenc} 
\usepackage[T1]{fontenc}    
\usepackage{hyperref}       
\usepackage{url}            
\usepackage{booktabs}       
\usepackage{amsfonts}       
\usepackage{nicefrac}       
\usepackage{microtype}      
\usepackage{xcolor}         
\usepackage{graphicx}

\DeclareUnicodeCharacter{0301}{~}

\usepackage[style=numeric,firstinits]{biblatex}
\title{Phenomenological classification of the Zwicky Transient Facility astronomical event alerts}

\author{%
  
  Dmitry A.~Duev\thanks{\url{https://duev.space}} \\
  Weights \& Biases, Inc.\\
  1479 Folsom St, San Francisco, CA 94103\\
  \\
  Division of Physics, Mathematics, and Astronomy\\
  California Institute of Technology\\
  Pasadena, CA 91125 \\
 
  \texttt{duev@caltech.edu} \\
   \And
   St\'{e}fan J. van der Walt \\
   Berkeley Institute for Data Science\\
   University of California, Berkeley\\
   Berkeley, CA 94720, USA \\
   \texttt{stefanv@berkeley.edu} \\
}

\begin{document}

\maketitle

\begin{abstract}
  The Zwicky Transient Facility (ZTF), a state-of-the-art optical robotic sky survey, registers on the order of a million transient events --- such as supernova explosions, changes in brightness of variable sources, or moving object detections --- every clear night, and generates associated real-time alerts. We present Alert-Classifying Artificial Intelligence (ACAI), an open-source deep-learning framework for the phenomenological classification of ZTF alerts. ACAI uses a set of five binary classifiers to characterize objects which, in combination with the auxiliary/contextual event information available from alert brokers, provides a powerful tool for alert stream filtering tailored to different science cases, including early identification of supernova-like and anomalous transient events. We report on the performance of ACAI during the first months of deployment in a production setting.
\end{abstract}

\section{Introduction}
\label{sec:intro}

Astronomical sky surveys observe a myriad of transient events in the dynamic sky originating from a wide range of astrophysical objects and processes. When detection of such events is performed in the image domain, an epochal image of a patch of the sky is compared to a reference image, usually adopting an image subtraction algorithm. 

Current and future large-scale astronomical sky surveys have the ability to detect millions of transient events, manifesting the need for automated separation of genuine astrophysical events from bogus detections and further categorization of real events. Events, real and bogus, are caused by a wide variety of phenomena, some of which are very hard to model.

To address this, sky surveys typically rely on binary real/bogus (RB) classifiers \cite{2007ApJ...665.1246B, 2002SPIE.4836...61A, bloom_towards_2008, brink_using_2012, wozniak_automated_2013, rebbapragada_time-domain_2015, goldstein2015des, wright2015panstarrs, 2017ApJ...836...97C, 2018arXiv180803626R, 2019MNRAS.489.3582D} or multi-class classifiers \cite{2020arXiv200803309C}.

\subsection{The Zwicky Transient Facility}

The Zwicky Transient Facility (ZTF)\footnote{\url{https://ztf.caltech.edu}} is a state-of-the-art robotic time-domain sky survey capable of visiting the entire visible sky north of $-30^\circ$ declination every night. ZTF observes the sky in the $g$, $r$, and $i$ bands at different cadences depending on the scientific program and sky region \cite{2019PASP..131a8002B, 2019PASP..131g8001G}. The 576 megapixel camera with a 47 deg$^2$ field of view, installed on the Samuel Oschin 48-inch (1.2-m) Schmidt Telescope, can scan more than 3750 deg$^2$ per hour, to a $5\sigma$ detection limit of 20.7 mag in the $r$ band with a 30-second exposure during new moon \cite{2019PASP..131a8003M, 2020PASP..132c8001D}.

The raw data are processed in real time at IPAC, the California Institute of Technology (Caltech). The ZTF Science Data System (ZSDS) housed at IPAC consists of data processing pipelines, data archives, infrastructure for long-term curation, and  services for data retrieval and visualization \cite{2019PASP..131a8003M}.

The part of ZSDS responsible for transient event detection and extraction performs image differencing of the science and reference images using the ZOGY algorithm \cite{2016ApJ...830...27Z}. If the resulting difference image is of sufficient quality, the pipeline then detects events from the point-source match-filtered images, performing detection on both the positive (science minus reference) and negative (reference minus science) images. Events are extracted with both aperture and point spread function (PSF) fit photometry, and additional source features are computed. The events are then lightly filtered to remove obvious false positives and image cutouts are generated \cite{2019PASP..131a8003M}. 

Events may have been triggered by a transient, a variable, or a moving object; or caused by, e.g., an optical or processing artifact. Metadata and contextual information, including image cutouts, are put into alert packets in the Apache AVRO format that are then distributed by the ZTF Alert Distribution System (ZADS) using Apache Kafka. The number of events detected nightly typically ranges from $10^5$ -- $10^6$.

\section{ACAI: a deep learning framework for the phenomenological classification of ZTF alerts}
\label{sec:acai}

In this work, we present Alert-Classifying Artificial Intelligence (ACAI), an open-source deep-learning framework for the classification of ZTF alerts\footnote{\url{https://github.com/dmitryduev/acai}}. ACAI uses a set of \textit{independent} binary classifiers to categorize objects into five phenomenological classes:

\begin{itemize}
    \item ``Hosted'' (\texttt{acai\_h}) -- genuine transients in the vicinity of a (host) galaxy with detectable morphology. Objects such as supernovae (SNe), novae, or cataclysmic variables occurring near galaxy-like objects get high \texttt{acai\_h} scores.
    \item ``Orphan'' (\texttt{acai\_o}) -- genuine orphan transients, i.e. when there are no identifiable ``hosts'' in their vicinity. This category catches asteroids and host-less (or with hosts that are too faint) transients.
    \item ``Nuclear'' (\texttt{acai\_n}) -- genuine transients occurring in the galaxy/quasar nucleus. Active galactic nuclei (AGN) and tidal disruption events (TDEs) fall into this category.
    \item ``Variable star'' (\texttt{acai\_v}) -- genuine brightness changes of variable stars detectable in the reference image or in a reference catalog.
    \item ``Bogus'' (\texttt{acai\_b}) -- bogus events such as cosmic ray hits, optical reflections, or data processing artifacts, get high \texttt{acai\_b} scores.
\end{itemize}

The main advantage of our approach with independent binary classifiers is significantly greater flexibility compared to typically used multi-class classifiers, where an object is assumed to have a single correct label of many, or multi-label classifiers, where a single system outputs probabilistic predictions of object class membership for multiple classes at once. 
If the performance on a particular class is deemed insufficient, retraining the classifier with new training data (or employing a different architecture) does not affect the system's performance on other classes. Adding new categories is straightforward and does not require rebuilding an entire multi-class or multi-label classifier.

Another advantage is more flexibility for the end-user. When used in the context of an alert broker, the ACAI scores can be mixed and matched with other criteria such as the information on cross-matches with external catalogs to create alert stream filters tailored to a particular science case.

Finally, our approach implicitly allows for anomaly detection. For example, when an object is classified as real but does not belong to any of the non-bogus categories with high confidence, or if it simultaneously receives high scores in multiple categories, it is potentially of interest.


\subsection{Data set}

In the case of alert classification, the training set must reflect the possible variations across different seeing conditions, filters, sky location, sensors, and include data artifacts caused by, for example, cross-talk or telescope reflections.

We have assembled a large training data set consisting of over 200,000 individual alerts -- about 70k alerts from known variable stars of different types, 35k from nuclear transients originating from known AGN and TDEs, 30k ``hosted'' transients from known supernovae and novae, 50k bogus alerts, and 30k orphan events, mostly from known Solar system objects.

We employed an active-learning-like approach of alternating between data labeling and classifier training with subsequent sampling of their predictions, both of confident ones and, more importantly, of those near the decision boundary and with the highest loss.

\subsection{Deep neural network architecture and training}

When building the classifiers, we had to explore a vast hyperparameter space. 

As input, our classifiers use 25 features from the ``candidate'' section of the alert packets\footnote{\url{https://zwickytransientfacility.github.io/ztf-avro-alert}} that include both derived characteristics/parameters of events as well as  contextual information, such as  cross-matches with external catalogs and full-sized (63x63 pixels) triplet (epochal science, reference, and difference) image cutouts.

\begin{figure*}
    \centering
    \includegraphics[width=1\textwidth]{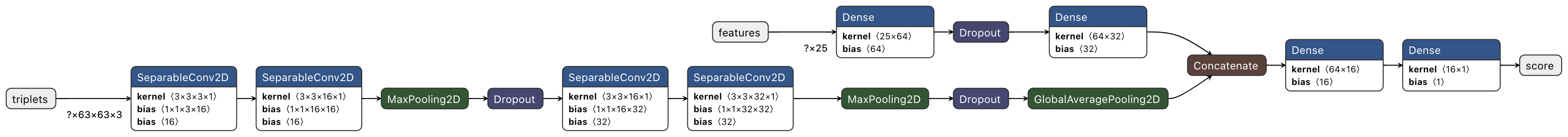}
    \caption{The architecture of the ACAI models.}
    \label{fig:architecture}
\end{figure*}

The classifiers were implemented using the \texttt{TensorFlow} library and its high-level \texttt{Keras} API \cite{tensorflow2015-whitepaper, chollet2015keras}. We employed the standard binary cross-entropy loss function, the Adam optimizer \cite{2014arXiv1412.6980K}, a batch size of 32, and a 81\%/9\%/10\% training/validation/test data split. Given the amount and diversity of gathered training data, we did not perform data augmentation. 

The input features were normalized; the same norms were used for all classifiers. We applied class balancing; the classifier performance was subsequently checked on the originally dropped ``negative'' examples and the small number of misclassifications (typically on the order of $1-2\%$) were added to the training set.
The training data were weighted per class. The class weights were further adjusted to balance precision (purity) and recall (completeness).
We used standard techniques to achieve high performance, such as learning rate reduction on a plateau and early stopping based on validation loss.

Initially we used a simple architecture that demonstrated satisfying performance, with a minimal, arbitrarily-chosen number of fully-connected and convolutional layers. As we expanded the input data sets, we ran several rounds of hyperparameter tuning using the Weights \& Biases' \texttt{sweeps}\footnote{\url{https://github.com/wandb/sweeps}} library \cite{wandb}. We tuned hyperparameters such as the number of fully-connected layers and neurons therein, number of convolutional filters and their sizes and types (regular or separable convolution), flattening versus global average pooling, dropout rates, activation functions, and the initial learning rate for the Adam optimizer. We also experimented with the inclusion/exclusion of particular alert features in/from the input.

We chose the simplest of the best-performing architectures (shown in Fig. \ref{fig:architecture}) and adopted it for all of the individual classifiers.

Training (including grid hyper-parameter search) was performed on a single Nvidia Tesla V100 16GB GPU on the Google Cloud Platform, and took just under 1 day.  

\section{ACAI performance}

\subsection{Test set performance}

\begin{figure*}
    \centering
    \includegraphics[width=0.8\textwidth]{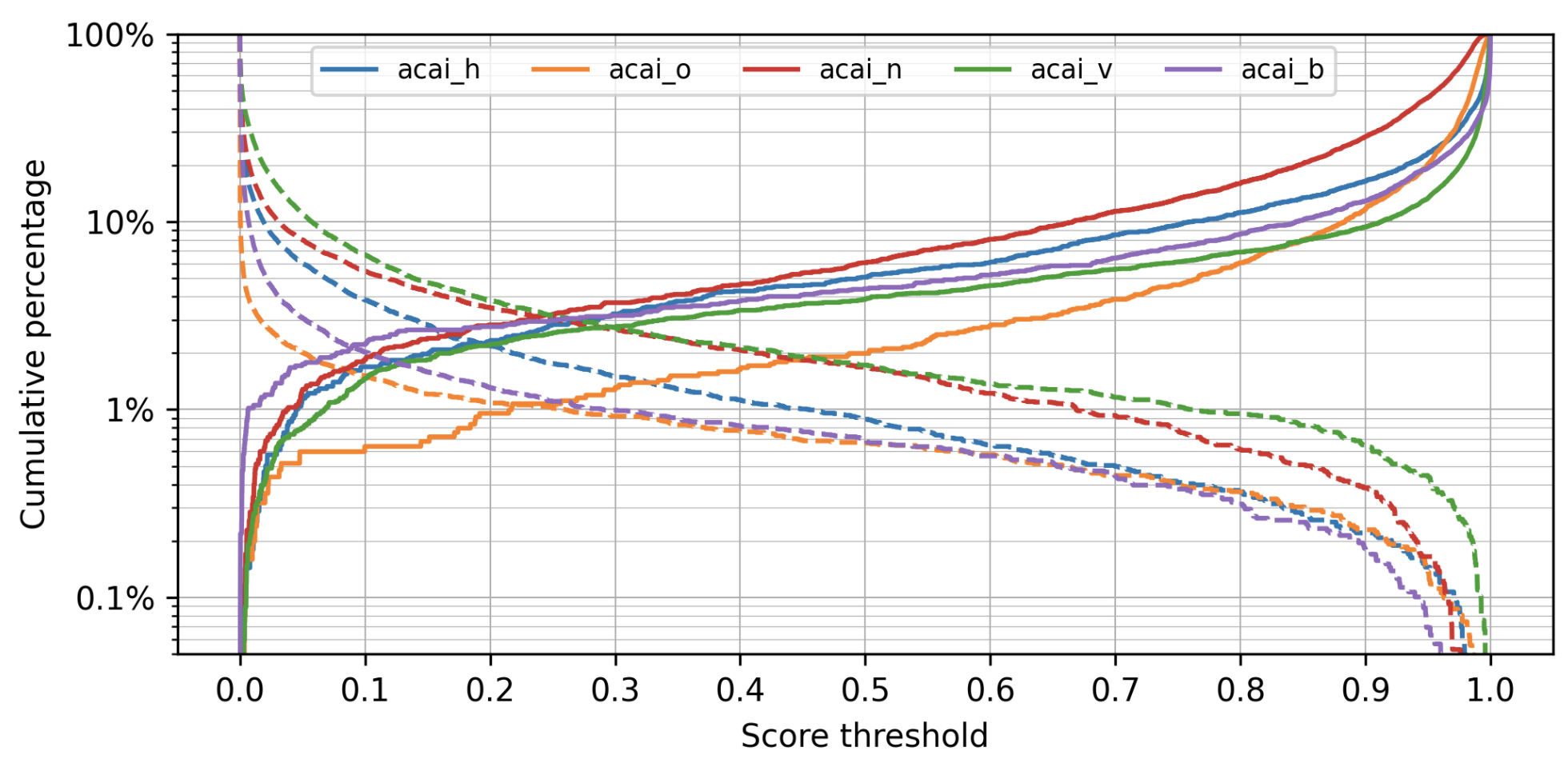}
    \caption{False negative (FNR; solid lines) and false positive rates (FPR; dashed lines) of the ACAI models as functions of the score threshold. The classifiers were evaluated on the test set consisting of $\sim20,000$ alerts.}
    \label{fig:fnr_fpr}
\end{figure*}

When evaluated on the test sets, all of the ACAI classifiers demonstrate 1-3\% balanced false positive / false negative rates for score thresholds around 0.2 (see Fig. \ref{fig:fnr_fpr}).

\subsection{Performance in production}

The ACAI classifiers have been deployed on Fritz\footnote{\url{https://github.com/fritz-marshal/fritz}}, an open-source data platform and alert broker used by the ZTF collaboration (consisting of hundreds of astronomers worldwide), in a production setting since early December 2020. When used in combination with auxiliary/contextual event information available from the Fritz's broker, Kowalski\footnote{\url{https://github.com/dmitryduev/kowalski}}, ACAI has proven to be a powerful tool (with configurable purity/completeness based on the score thresholds) for alert stream filtering tailored to different science cases, yielding numerous discoveries. These include early detection of SN-like events (often after only a single alert), identification of tidal disruption events, identification of unknown Solar system objects, and more. Finally, we have been able to identify multiple astrophysically-interesting anomalous objects, detailed investigation of which is currently underway.



\subsubsection{Early detection of ``hosted'' SNe}

As an demonstration, we have been running a simple alert stream filter that selects events with high $acai\_h$ scores and low scores from other ACAI classifiers, detected fewer than 5 times as a positive source in the difference image, and not originating from a know Solar system object. This can be expressed in pseudo code as follows:

\begin{verbatim}
acai_h > 0.8 AND MAX(acai_b, acai_n, acai_v, acai_o) < 0.1
  AND n_detections < 5 AND is_positive_subtraction AND NOT is_known_ss_object
\end{verbatim}

From January through September 2021, this filter has yielded 12,035 objects, of which only 52 have been marked as likely bogus ($0.4\%$ FPR). The majority of the selected objects exhibit SN-like light curves, at peak fainter than $\sim 19.5-20$ mag. 






\section{Limitations}

We note that the training sets for the different classifiers overlap, so while the classifiers are executed independently, their output is expected to be correlated.
Furthermore, ACAI is designed to aid in (early) discovery of interesting events and is intended to be used alongside other information. More work is required to characterize completeness of samples selected with the use of ACAI.

\section{Conclusions}
We have demonstrated that, by putting together a large, representative, and uncontaminated data set with a relatively simple deep model, we can achieve excellent classification performance. To improve it even further, we will retrain and deploy new classifiers as more labeled data are collected. Our setup will be useful for other surveys including the Vera Rubin Observatory's Legacy Survey of Space and Time (LSST, \cite{ivezic2008lsst}) in the near future.

\begin{ack}
D. A. Duev would like to thank Ivan Duev for assistance with data labeling.

D. A. Duev acknowledges support from Google Cloud and from the Heising-Simons Foundation under Grant No. 12540303. 

Based on observations obtained with the Samuel Oschin Telescope 48-inch and the 60-inch Telescope at the Palomar Observatory as part of the Zwicky Transient Facility project. ZTF is supported by the National Science Foundation under Grant No. AST-1440341 and a collaboration including Caltech, IPAC, the Weizmann Institute for Science, the Oskar Klein Center at Stockholm University, the University of Maryland, the University of Washington, Deutsches Elektronen-Synchrotron and Humboldt University, Los Alamos National Laboratories, the TANGO Consortium of Taiwan, the University of Wisconsin at Milwaukee, and Lawrence Berkeley National Laboratories. Operations are conducted by COO, IPAC, and UW.
\end{ack}

\small
\printbibliography

\end{document}